\documentclass[preprint,prd,aps,showkeys]{revtex4-1}
\usepackage{amsmath}
\usepackage{natbib}
\usepackage{graphicx}

\begin{document}
\title{New leaves of the tree: percolation analysis for cosmic web with discrete points}
\author{Jiajun Zhang}
\affiliation{School of Physics and Astronomy, Shanghai Jiao Tong University, Shanghai, China}
\affiliation{Department of Physics, The Chinese University of Hong Kong, Shatin, N.T., Hong Kong}
\author{Dalong Cheng}
\affiliation{School of Physics and Astronomy, Shanghai Jiao Tong University, Shanghai, China}
\author{Ming-Chung CHU}
\affiliation{Department of Physics, The Chinese University of Hong Kong, Shatin, N.T., Hong Kong}

\begin{abstract}
Percolation analysis has long been used to quantify the connectivity of the cosmic web. Most of the previous work is based on density fields on grids. By smoothing into fields, we lose information about galaxy properties like shape or luminosity. The lack of mathematical modelling also limits our understanding for the percolation analysis. In order to overcome these difficulties, we have studied percolation analysis based on discrete points. Using a Friends-of-Friends (FoF) algorithm, we generate the $S-bb$ relation, between the fractional mass of the largest connected group ($S$) and the FoF linking length ($bb$). We propose a new model, the Probability Cloud Cluster Expansion Theory (PCCET) to relate the $S-bb$ relation with correlation functions. We show that the $S-bb$ relation reflects a combination of all orders of correlation functions. Using N-body simulation, we find that the $S-bb$ relation is robust against redshift distortion and incompleteness in observation. From the Bolshoi simulation, with Halo Abundance Matching (HAM), we have generated a mock galaxy catalogue. Good matching of the projected two-point correlation function with observation is confirmed. However, comparing the mock catalogue with the latest galaxy catalogue from SDSS DR12, we have found significant differences in their $S-bb$ relations. This indicates that the mock galaxy catalogue cannot accurately retain higher order correlation functions than the two-point correlation function, which reveals the limit of the HAM method. As a new measurement, $S-bb$ relation is applicable to a wide range of data types, fast to compute, robust against redshift distortion and incompleteness, and it contains information of all orders of correlation functions.   
\end{abstract}

\maketitle
\section{Introduction}

Cellular or clumpy? That is the question. In the last century, when the first generation of redshift surveys were available, people were confused by the filamentary structure of galaxy distribution \citep{CfA1982,CfA1989}. The existence of the cosmic web was qualitatively confirmed, but a quantitative description of it was not available until Zeldovich's percolation analysis \citep{Zel1982,ZES1982,Shan1983}. We classify the percolation analyse for the cosmic web into two branches. The first one, dealing with discrete points, is called continuum percolation, whereas the second one, dealing with the density field on grids, is called site percolation. Continuum percolation analysis was also called cluster analysis in 1980s, because the same technique was used to define galaxy clusters. Cluster analysis can clearly tell the difference between a clumpy structure and web-like pattern \citep{PerSim1984}. However, Dekel \& West \citep{PerSim1985dead} concluded that cluster analysis was insensitive to cosmological parameters, and it was also non-trivially affected by volume and point density. Thereafter, the site percolation analysis became popular, and properties of site percolation in large scale density fields have been studied in detail \citep{PerSim1993,PerSim1995,PerSim1996,PerSim2000,PerSim2004,PerSim2005,PerSim2009,PerSim2011,PerSim2012} . By smoothing points into field, the technique has also been used to look for properties of galaxy distribution \citep{PerObs1989,PerObs1997,PerObs2005}. We would like to point out that direct comparison between the percolation properties of density field and galaxy distribution is not fair, because galaxies don't trace the density field perfectly. If we can generate a mock catalogue of galaxies and compare it with a real redshift survey, the comparison will be fair and the result will give us some useful insights in properties of the cosmic web. 

On the other hand, the establishment of the $\Lambda$CDM model and rapid development of N-body simulation techniques provide us a general picture of the formation of the cosmic web \citep{CosmicWeb1,CosmicWeb2,CosmicWeb3}. We are now confident about the success of the $\Lambda$CDM model in large scales. However the three-point, four-point or even N-point correlation functions should also be tested \citep{3pcf2004,3pcf2005,3pcf2007,3pcf2015,npcf}, though they are all very difficult to calculate and  model in theory. In principle, the two-point correlation function costs $O(N^{2})$ calculations, and the three-point correlation function costs $O(N^{3})$ calculations, so on and so forth. Although we have better ways to optimize the calculation, higher order correlation functions  still require unbearable calculation resources. Thus, several methods to characterize the cosmic web have been proposed, such as Minkowski Functional, Minimum Spanning Tree, and Genus Analysis \citep{MinFun1,MinFun2,MST2007,Genus2005}. These methods focus on the morphology and topology of the cosmic web. 

The comparison between simulation and observation is not straight forward. We are still lacking of knowledge about galaxy formation and evolution, and matching galaxies with simulated dark halos heavily depends on model \citep{GalFormation1,GalFormation2,GalFormation3,GalFormation4}. Surprisingly, Halo Abundance Matching (HAM), a newly developed technique based on simple arguments, generates mock galaxy catalogues successfully \citep{HAMmainref,HAMref1,HAMref2}. HAM follows the  halo mass, halo circular velocity, halo potential well, galaxy circular velocity, galaxy luminosity, galaxy stellar mass, independent of galaxy formation model, and HAM is easy to understand.     

Shandarin et al. \citep{NaturevsNurture2010} developed and studied the percolation analysis, treating voids and superclusters on an equal footing and finding the nature and nurture of the formation of cosmic web. Using a similar method, the SDSS galaxy sample has been tested with the percolation analysis \citep{PerObs2005}. However we are still far from understanding the percolation transition in
the cosmic web. Density estimation is a prior step for site percolation analysis, but estimating  the density field from galaxy observation is not easy. We should take into account galaxies' different properties in the percolation analysis.  

Based on these considerations, we used the continuum percolation analysis to calculate the $S-bb$ relation, where $bb$ stands for the FoF linking length and $S$ stands for the mass fraction of the largest group for the given linking length $bb$ \citep{FoFref}. It is quite similar to the largest cluster statistics in previous percolation analyses \citep{PerSim1985dead}. However, we find that the continuum percolation analysis can distinguish among cosmological models with high resolution simulations, whose resolution was not available in 1985. We propose the Probability Could Cluster Expansion Theory (PCCET), to calculate the $S-bb$ relation theoretically. We measure the $S-bb$ relation in our mock galaxy catalogue and compare it with a real galaxy catalogue. We find that HAM cannot reproduce the $S-bb$ relation as well as the projected two-point correlation function. 

\section{Methodology}
\subsection{S-bb Relation and Percolation Transition}
\begin{figure}
\includegraphics[width=\textwidth]{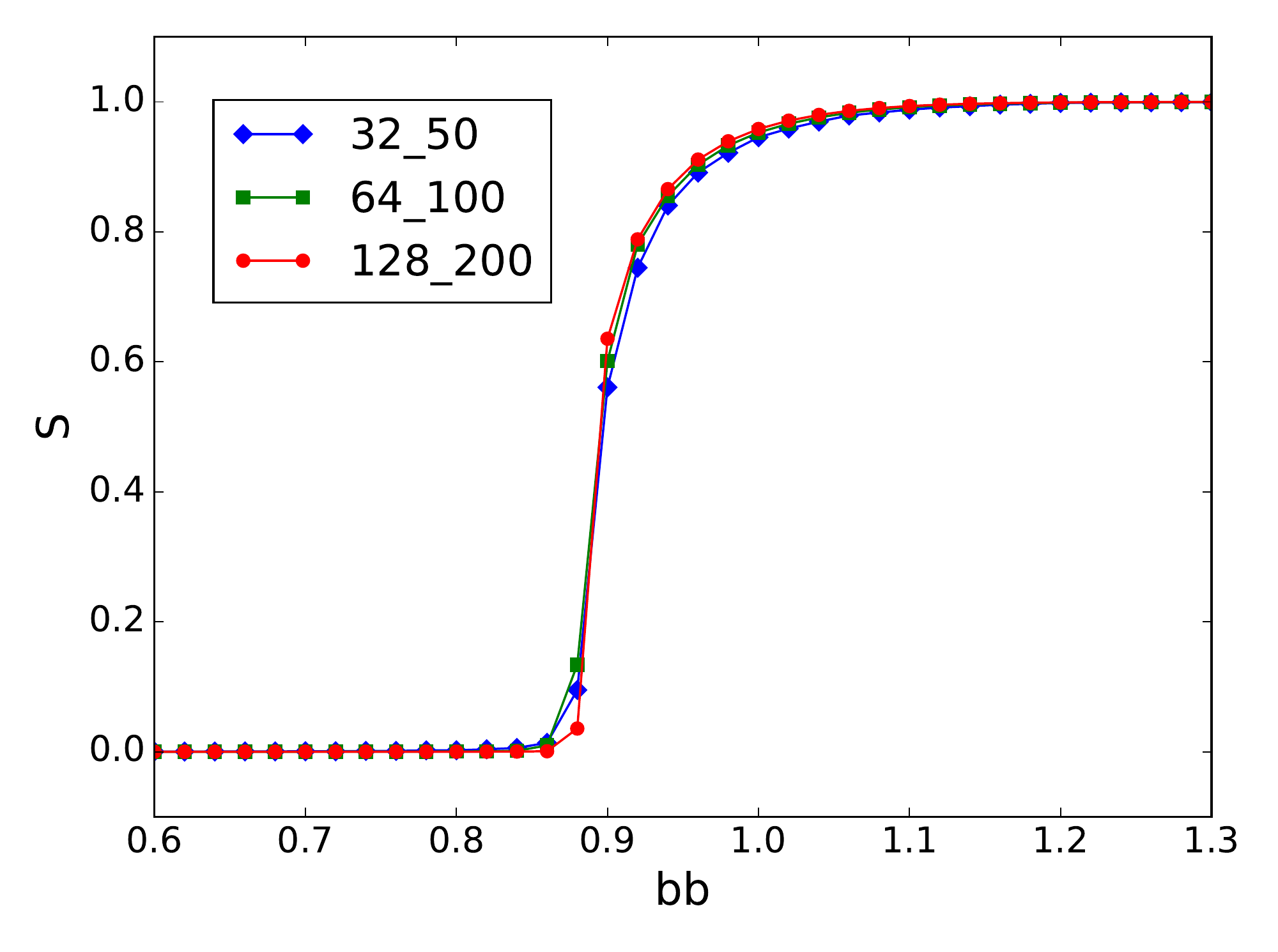}
\caption{Three sets of $S-bb$ relation for random distributions are shown here, corresponding to $32^{3}$, $64^{3}$ and $128^{3}$ particles in boxes with $50h^{-1}$Mpc, $100h^{-1}$Mpc, and $200h^{-1}$Mpc sides. Note that these three curves are degenerate and the transition threshold is around $bb=0.88$.}\label{random}
\end{figure}

For any discrete point distribution, we can always apply the FoF algorithm to find a group of points. The points can be massive points in N-body simulations, halos in simulated halo catalogues or galaxy catalogues in redshift space. We define $S_{1}$ and $S_{2}$ as:

\begin{equation}
   S_{1}(bb)=N_{largest}/N_{total},
\end{equation}
\begin{equation}
   S_{2}(bb)=N_{second-largest}/N_{total},
\end{equation}
 where
 $N_{total}$ is the total number of points,
 $N_{largest}$ ($N_{second-largest}$) is the number of points in the largest (second largest) group found by FoF, and normally, we refer to the $S-bb$ relation the function of $S_{1}(bb)$.
 It is obvious that $S_{1}(bb)$ reveals quite similar information as the cluster analysis and site percolation analysis.
 Our simulation tests are all done using \texttt{Gadget2} \citep{gadget2}. Fig.~\ref{random} is the $S-bb$ relation for a random distribution as a comparison for later figures. It is not surprising that all three sets of data corresponding to different box sizes and number of particles give the same $S_{1}(bb)$ and the transition threshold is around $bb=0.88$.

We define the transition threshold as the $bb$ where $S_{2}(bb)$ gets the maximum value. Another definition of the transition threshold was also proposed in \citep{PerSim1993}; in our case, it can be defined as the $bb$ where the mass-weighted sum of the number of connected structures, $\mu^2=\dfrac{\Sigma_n N_n}{N_{groups}}$, where $N_n$ is the number of particles in the $n^{th}$ group and $N_{groups}$ is the total number of connected structures, gets the maximum value. 

\begin{figure}
\includegraphics[width=\textwidth]{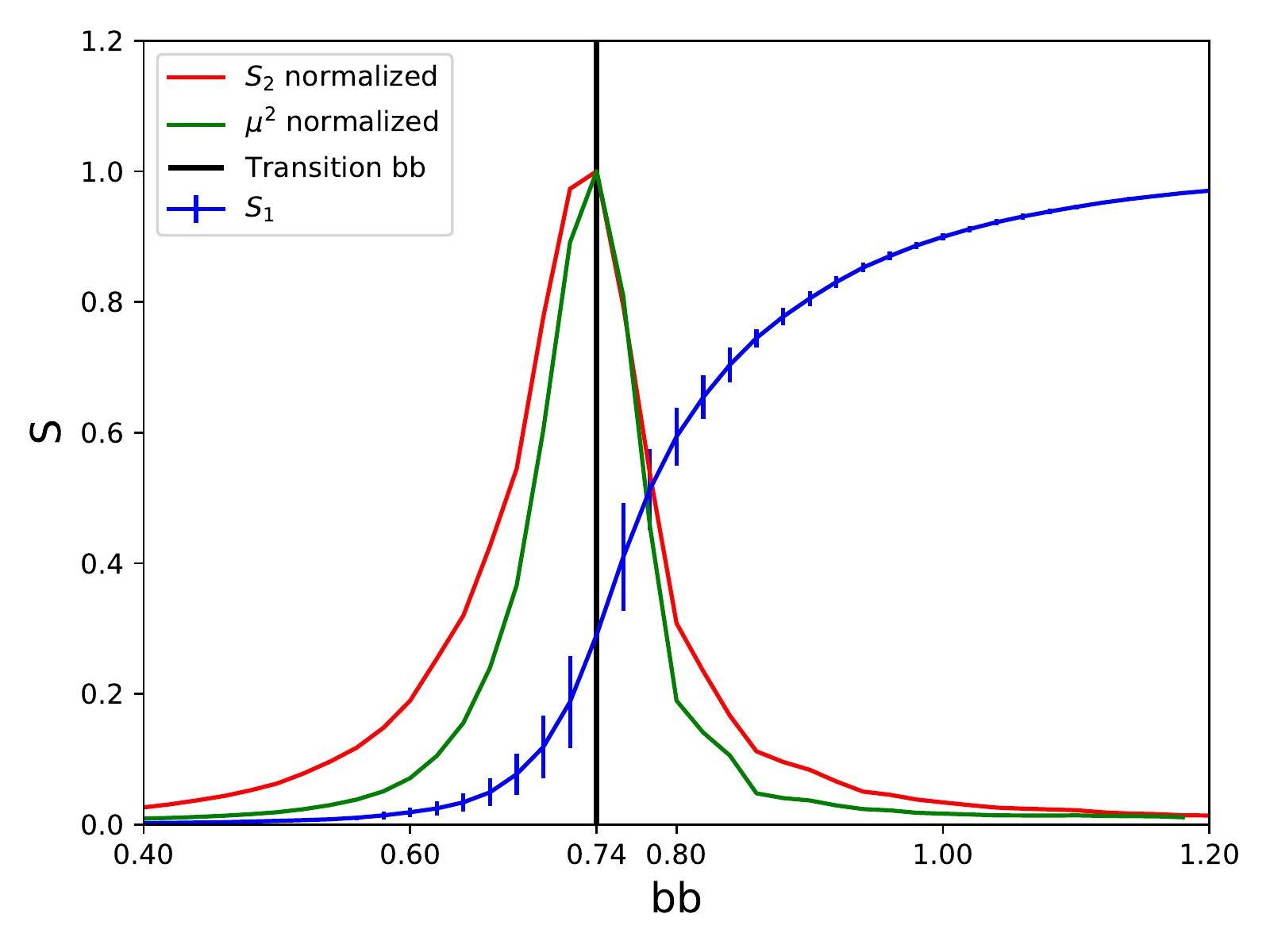}
\caption{The $S-bb$ relation measured from the MDR1 simulation \citep{multidark1}. The blue line is $S_{1}$, the red line is the normalized $S_{2}$, and the green line is the normalized $\mu^2$. The transition threshold is shown as the vertical black line, at $bb=0.74$ in this plot. The sharp peak of $S_{2}$ reveals the nature of percolation, that most massive structures combine into the largest structure which show up beyond the size of superclusters. The FWHM of $\mu^2$ and $S_2$ are similar, but the FWHM of $\mu^2$ is $\sim 10\%$ smaller than that of $S_2$. The shape of $S_{1}$ is very much different from the random sample shown in Fig.~\ref{random}. It shows clearly that the cosmic web is not random.}\label{transition_def}
\end{figure}

We measured the $S_1$, $S_2$ and $\mu^2$ using the halo catalogue of the MDR1 simulation\citep{multidark1} with haloes with mass larger than $10^{12}M_{\odot}$. The MDR1 simulation was performed in a box with $1h^{-1}$Gpc sides, and so we separated the sample into 64 subsamples of boxes with $250h^{-1}$Mpc sides to estimate the cosmic variance, shown as the errorbars in Fig.~\ref{transition_def}. The transition threshold $bb_c$ is 0.74 and the power index $\alpha$ given by $S_1=(bb-bb_c)^{\alpha}$ was measured to be $\alpha=0.46$, which is consistent with \citep{PerSim1993}. 
 
As shown in Fig.~\ref{transition_def}, both definitions of the transition threshold are good, and their FWHM are similar. We will discuss the meaning of the transition threshold in the definition of $S_2$ later for easier understanding. When the percolation transition happens, the largest structure grows tremendously as $bb$ gets larger, and the second largest structure reaches its most massive point. After that, the mass of the second largest structure becomes smaller, not because the structure shrinks, but because the original second largest structure is connected to the largest one while another structure is recognized as the new second largest one. This connection and replacement reveals the nature of percolation: the largest structure becomes the one and the only one dominant structure, connecting most of the high density regions together and leaving only the far apart particles in voids alone.  
Thus, the definition of transition threshold based on either $S_2$ or $\mu^2$ is appropriate.

While the $\Lambda$CDM (cold dark matter) model is not clearly distinguishable from the $\Lambda$WDM (warm dark matter) model, the $\Lambda$HDM (hot dark matter) model is clearly ruled out by observation of large scale structures. We would like to test our percolation analysis with these three models. The $S-bb$ relations are expected to be indistinguishable for $\Lambda$CDM and $\Lambda$WDM models, but those for $\Lambda$HDM and $\Lambda$CDM models should be clearly different. This test is the bench mark test of percolation analysis.   We generate the initial conditions with \texttt{2LPTic} \citep{2lptic} and run the simulation from redshift 49 to redshift zero using the N-body simulation code \texttt{Gadget2}\citep{gadget2}. The transfer function and additional thermal velocity for HDM and WDM are given by \citep{WDMmodel}. These three simulations share the same realization. Some basic parameters for the simulations are given in Table.~\ref{HWC}. 
\begin{table}[h!]
\centering
\caption{Basic parameters for three simulations. EH is the power spectrum given by \citep{EisensteinHu}.}
\begin{tabular}{c c c c c}
\hline
Model & $m$(keV) & IC & N & $L$($h^{-1}$Mpc)\\
\hline
CDM & $10^{5}$ & EH & $256^{3}$ & 250\\
WDM & 1 & EH & $256^{3}$ & 250\\
HDM & 0.05 & EH & $256^{3}$ & 250\\
\hline
\end{tabular}
\label{HWC}
\end{table}
\begin{figure}
\includegraphics[width=0.45\textwidth]{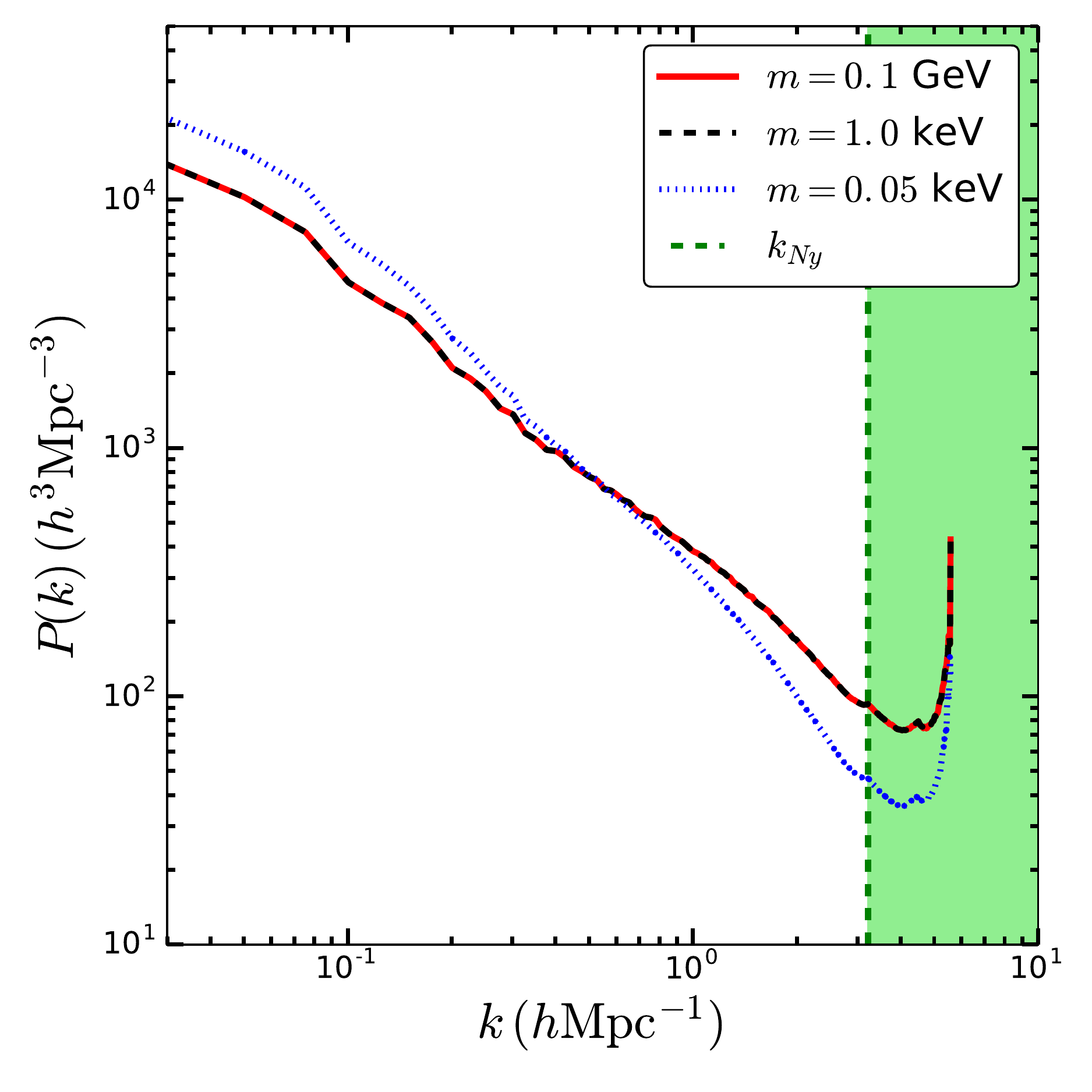}
\includegraphics[width=0.45\textwidth]{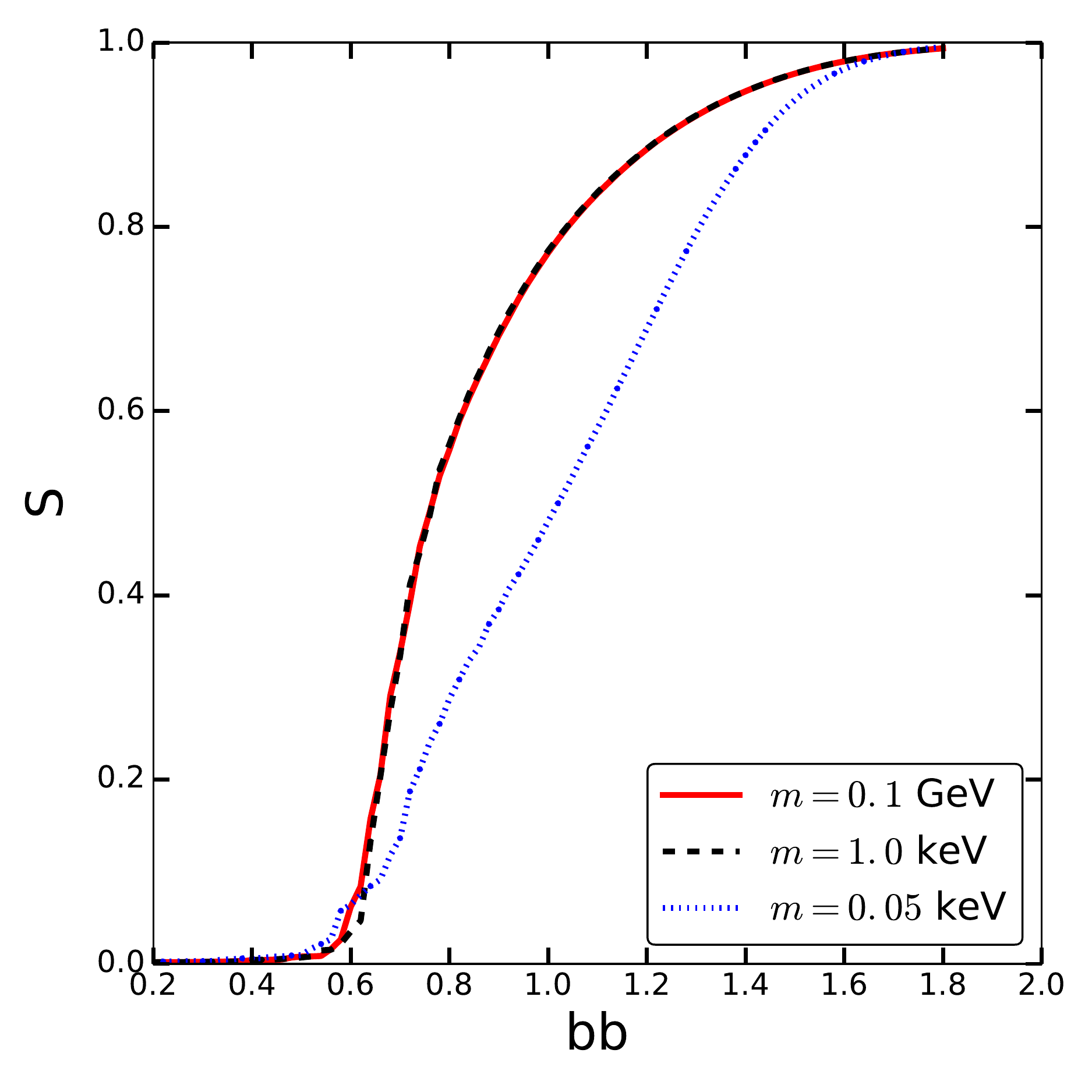}
\caption{Comparison of the matter power spectra and $S-bb$ relations for Hot (red curves), Warm (green curves) and Cold (blue curves) Dark Matter models. The green shaded region on the left panel indicates the resolution limit of the simulations.}\label{HWCsbb}
\end{figure}
As shown in Fig.~\ref{HWCsbb}, the HDM model is quite different from CDM and WDM models in their matter 
power spectra and $S-bb$ relations. The matter power spectrum is calculated by the publicly available
code \texttt{ComputePk}\citep{computepk}. It is well accepted that the HDM model has been ruled out by
measuring the power spectrum in both CMB \citep{ade2016planck} and large scale structures \citep{SDSSDR12}.
We have shown here that the difference between HDM model and CDM model is significant enough
in the $S-bb$ relation to be used as a supplementary. On the other hand, the unique
advantage of using the $S-bb$ relation will be discussed in the following sections. 

\subsection{Probability Cloud Cluster Expansion Theory}
The percolation analysis has been used for a long time to study the cosmic web. For site percolation there are analogies to similar questions in many other fields \citep{PercolationTheory,PercolationTheory2,PercolationTheory3,PercolationTheory4}. But for discrete points, there is still no available theory to link percolation with basic cosmological parameters. Here we propose the Probability Cloud Cluster Expansion Theory (PCCET) to connect the $S-bb$ relation with correlation functions. The idea is as follows:

\begin{enumerate}
\item Consider a box with only one particle and take it as the only member in the largest group. So the $S-bb$ relation will be a constant line $S=1$.
\item Add in another particle randomly, the probability of linking these two particles in a single group $P(bb)$ being monotonically related to the linking length $bb$. The $S-bb$ relation will be a step function jumping from $S_{1}=0.5$ to $S_{1}=1.0$ at a certain transition value of $P$: $S=\frac{1}{2}+\frac{1}{2}P$.
\item Add in the third particle randomly, the probability of linking any two of them being still $P$ and the probability of linking all three of them shall be $P^{2}$. While from a certain particle as the starting point of the group, the probability of linking the other two particles shall be $2P$ but deducting $P^{2}$ to avoid over counting. The $S-bb$ relation will then be a three-step function, with steps at $1/3$, $2/3$, and $1$: $S=\frac{1}{3}+\frac{1}{3}P+\frac{1}{3}P(2P-P^{2})$.
\item Continuing on, we consider more and more particles, taking care of over counting and over deducting, and we finally find the relation $S(P)$ when the number of particles goes to infinity. We hide all information about $bb$ in the function $P(bb)$.  
\end{enumerate}
So in general, we can write $S(P)$ as

\begin{equation}\label{eq:S-P}
\begin{cases}
a_{1}=\frac{1}{N}, \\
a_{n}=a_{n-1}b_{n-1} (n>1), \\
b_{n}=1-(1-P)^{n}, \\
S=\sum\limits_{n=1}^N a_{n}.
\end{cases}
\end{equation}

This series expansion is the reason why we call it cluster expansion theory. Cluster expansion is often used in condensed matter physics. It can successfully account for phase transitions in some typical problems. We borrow the idea of cluster expansion because of the similarity between percolation transition and phase transition. We have shown that $S(P)$ is monotonically increasing with $P$ and has a limit between 0 and 1. When $N\rightarrow\infty$, $S(P)$ will become the Euler Function. The proof in detail is given in Sec.~\ref{sec:SbbApp}. 

In the framework of this theory, a particle cannot be treated as a discrete point. In fact, it is better to take the particle as a representation of the probability cloud. For a random distribution, the probability for a particle to be closer to another particle than distance $r$ is inversely proportional to the volume of a sphere with radius $r$. However, we need a volume to set as the standard for this probability. Here we take all the particles in the theory as test points. We can first put in the test center. We imagine that every test center has its own cube with periodic boundary conditions. The length of each side of this test cube can be set as the average distance between particles in a real simulation. Then the distance to the center can be simply understood as the linking length $bb$. If $bb$ is larger than $1/2$, the influence of this test center will reach outside the test box. In such a case, we shall remove the influence region outside the test box and only consider the volume inside the box as the real probability distribution region. Shown in Eq.~\ref{eq:P-bb} is the $P-bb$ relation for a random distribution.

\begin{table}
\begin{equation}\label{eq:P-bb}
P=
\begin{cases}
\frac{4}{3}\pi(\frac{bb}{2})^{3},& 0\leq bb\leq1,\\
\frac{4}{3}\pi(\frac{bb}{2})^{3}-6\pi(\frac{bb-1}{2})^{2}(\frac{bb}{2}-\frac{bb-1}{6}),& 1<bb\leq\sqrt{2},\\
1-\int_{0}^{1-\sqrt{bb^{2}-2}}dx\int_{0}^{\sqrt{bb^{2}-1-(x-1)^{2}}}dy\int_{0}^{1-\sqrt{bb^{2}-(x-1)^{2}-(y-1)^{2}}}dz,& \sqrt{2}<bb\leq\sqrt{3},\\
1,& bb>\sqrt{3}. 
\end{cases}
\end{equation}
\end{table}

\begin{figure}
\includegraphics[width=0.45\textwidth]{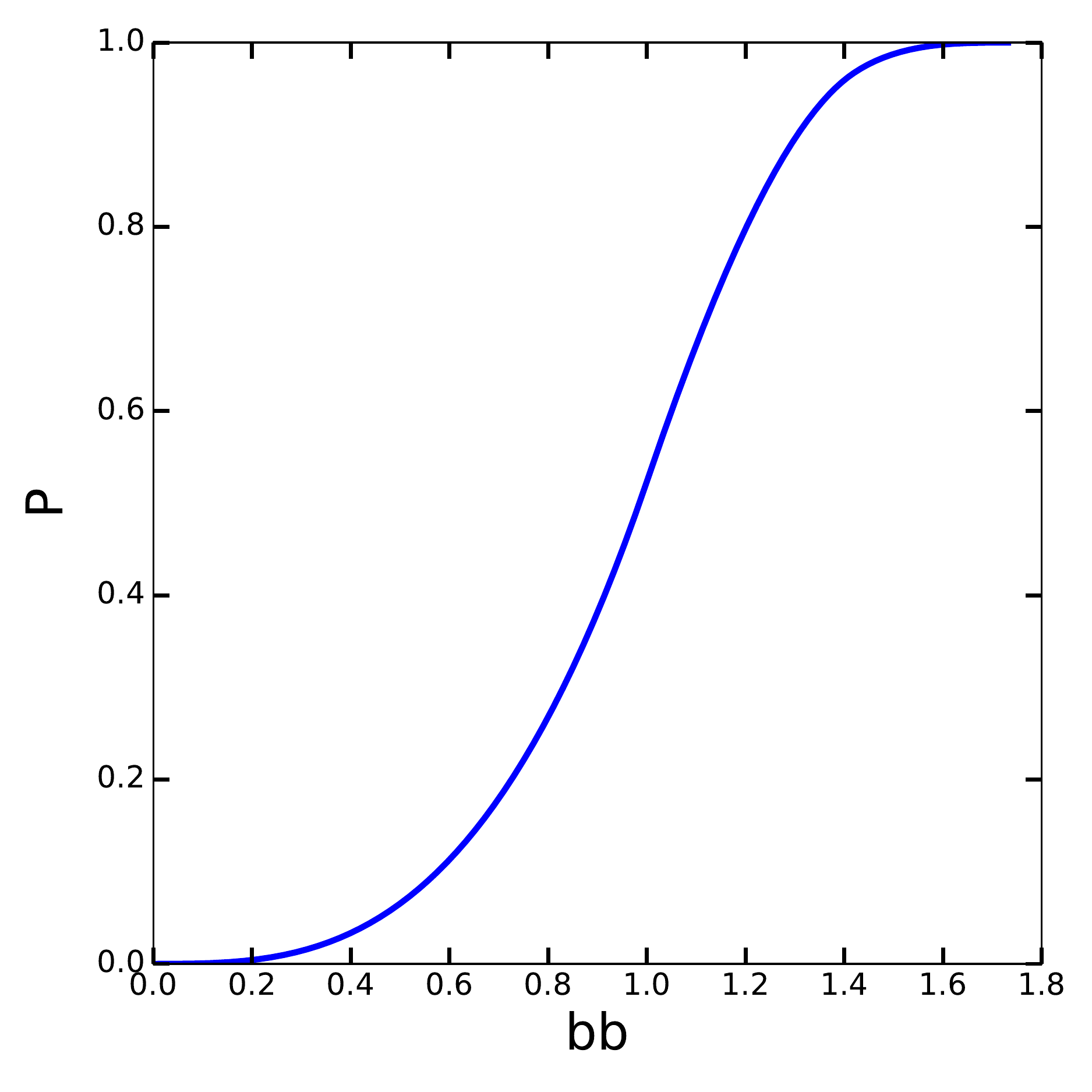}
\includegraphics[width=0.45\textwidth]{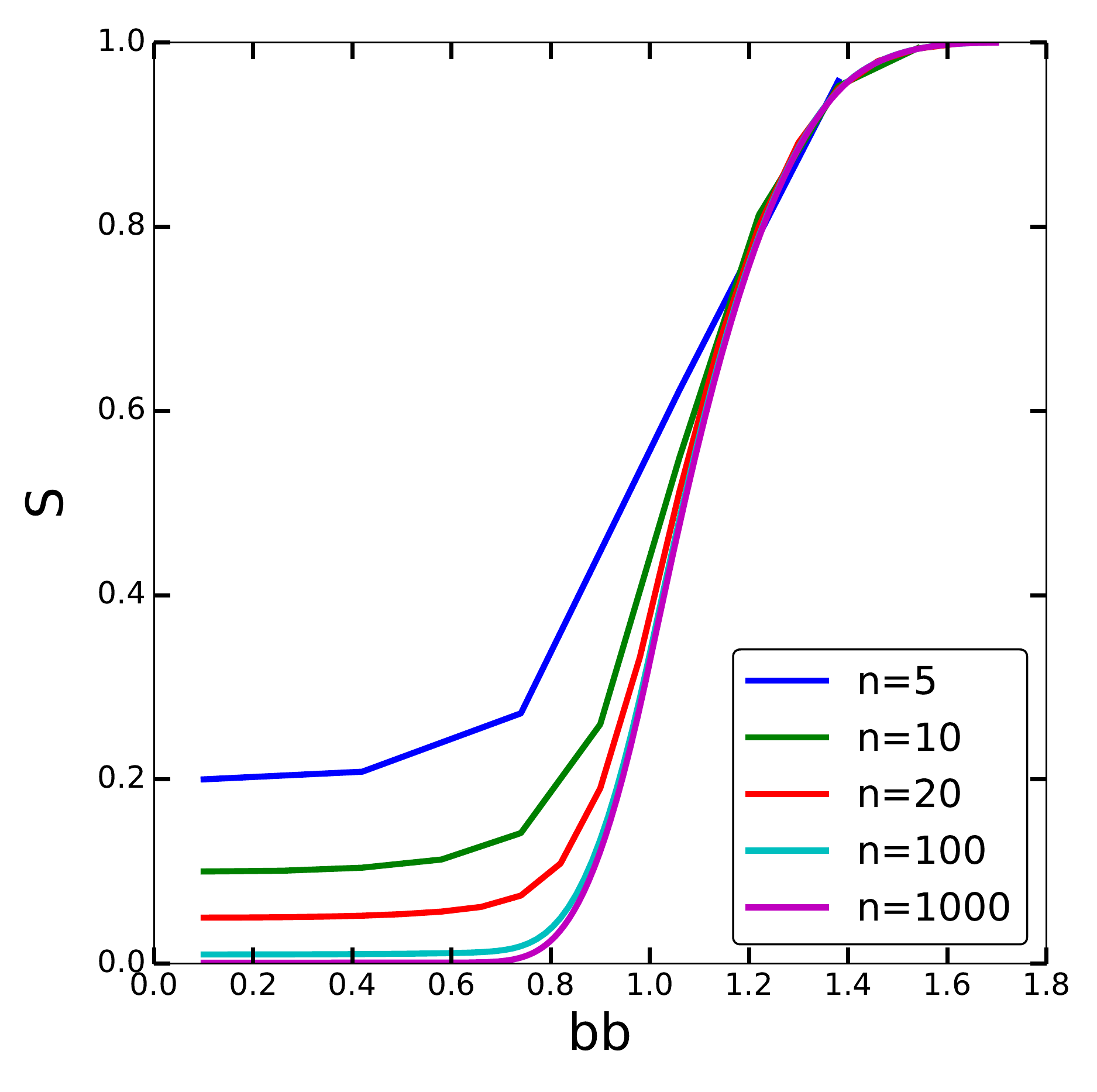}
\caption{The $P-bb$ relation given in Eq.~\ref{eq:P-bb} is shown on the left. The $S-bb$ relation according to Eq.s~\eqref{eq:P-bb} and ~\eqref{eq:S-P} is shown on the right. Notice the differences of the curves with different orders of expansion $n$ given by Eq.~\eqref{eq:S-P}.}\label{p-bb-s-bb}
\end{figure}

From the PCCET, we can find the relation between percolation and correlation function. Because of the cluster expansion, the 2-point correlation function is the correction to $P$ as the probability of linking up 2 points, while the 3-point correlation function is the correction to $P^2$ as the probability of linking up 2 pairs of points, etc. In general, the n-point correlation function is the correction to $P^{(n-1)}$ as the probability of linking up $n-1$ pairs of points. This shows that percolation actually contains the information of all orders of correlation functions. Our probability cloud cluster expansion theory also shows how every order of correlation function contributes to the $S-bb$ relation, which represents the percolation phenomenon in the cosmic web.

Unfortunately, in the non-linear regime, high order correlation functions are also quite large and not necessarily smaller than the 2-point correlation function. So based on our theory, it is still not possible to calculate the whole $S-bb$ relation analytically. It is important to use N-body simulations to set up constraints for the $S-bb$ relation. After all, PCCET is still the only model explaining percolation. It can be used to find the $S-bb$ relation in the linear-regime and it is also possible that in the future, we may find a way to use it in the non-linear regime to extract high order correlation functions.

\subsection{Halo Abundance Matching}
From PCCET, we can see that the $S-bb$ relation contains information of all orders of correlation functions. Thus we would like to see whether it provides new information in comparing simulation with observation. We first need to generate a mock galaxy catalogue and check the projected 2-point correlation function to see whether it matches the real galaxy catalogue. Then we will compare the $S-bb$ relations of the mock and real galaxy catalogues.

Based on the Bolshoi simulation \citep{Bolshoi1,Bolshoi2} BDM halo merger tree, we follow the work in \citep{HAMmainref} using Halo Abundance Matching to generate a mock galaxy catalogue. Here is the process:
\begin{enumerate}
\item We take the Bolshoi simulation BDM halo catalogue at redshift 0 to extract the final positions and peculiar velocities of the galaxies. We assume that every BDM halo, including the host haloes and subhaloes, can host at most one galaxy at its center. 
\item From the merger tree of BDM haloes in the Bolshoi simulation, we find the maximum circular velocity $V_{max}$ in the merger history with the halo circular velocity defined as the maximum value of the circular orbital velocity according to the density distribution of the halo.
\item We adopt the luminosity function from \citep{LumFun}, which is of the Schechter form. The luminosity function is:
\\
\begin{equation}\label{lumfun}
 \begin{split}
 \Phi(M)dM=0.4\log_{10}\Phi_{*}10^{-0.4(M-M_{*})(\alpha+1)}
 \\ 
 \times \exp(-10^{-0.4(M-M_{*})})dM, 
 \end{split}
\end{equation}

where $M$ is the absolute magnitude and $\Phi_{*}=0.0078$, $M_{*}-5\log_{10}h=-20.83$, $\alpha=-1.24$, given by \citep{LumFun} using SDSS6 data.
\item We sort the haloes in the Bolshoi simulation according to their $V_{max}$. Record every halo's ranking. Thus, we can easily obtain the number density of haloes with $V_{max}$ larger than a certain value by the ranking.
\item The integral of the luminosity function is an incomplete gamma function. We apply an one-one mapping from the number density of haloes with $V_{max}$ larger than a certain value $V$ to the integral of the luminosity function. Thus we obtain an one-one mapping between $V$ and absolute magnitude $M$.
\item Combining the data together, we get the halo catalogue with the galaxy absolute magnitude. This is also the mock galaxy catalogue with scale $250h^{-1}$Mpc. In this catalogue, we have positions, peculiar velocities and luminosities for the mock galaxies at redshift 0.
\end{enumerate}

Halo Abundance Matching has been proven quite successful in recreating a galaxy catalogue which can reproduce the observed projected galaxy 2-point correlation function in the bins of absolute magnitude close to the steep transform range in the luminosity function, around $\Phi=\Phi_{*}$. Many works follow the idea of HAM \citep{HAMref1,HAMref2}. There are slightly different methods to obtain a mock catalogue based on the same basic idea of HAM. Comparing to \citep{HAMmainref}, we did not take stochastic scattering into account. Since there are obvious bias to dimmer galaxies in stochastic scattering, we avoid this process. On the other hand, we take maximum circular velocity in the merger history of every halo as its monotonic mapping label. The physical insight of this picture is clear. Gas should fall into the center of a halo much faster than the merger rate, and once the gas has fallen into the halo center, it is so dense that the merger process cannot do much to the central dense region. So the galaxy luminosity depends more on the steepest potential well in the merger history, denoted by the maximum circular velocity $V_{max}$.

\begin{figure}
\includegraphics[width=\textwidth]{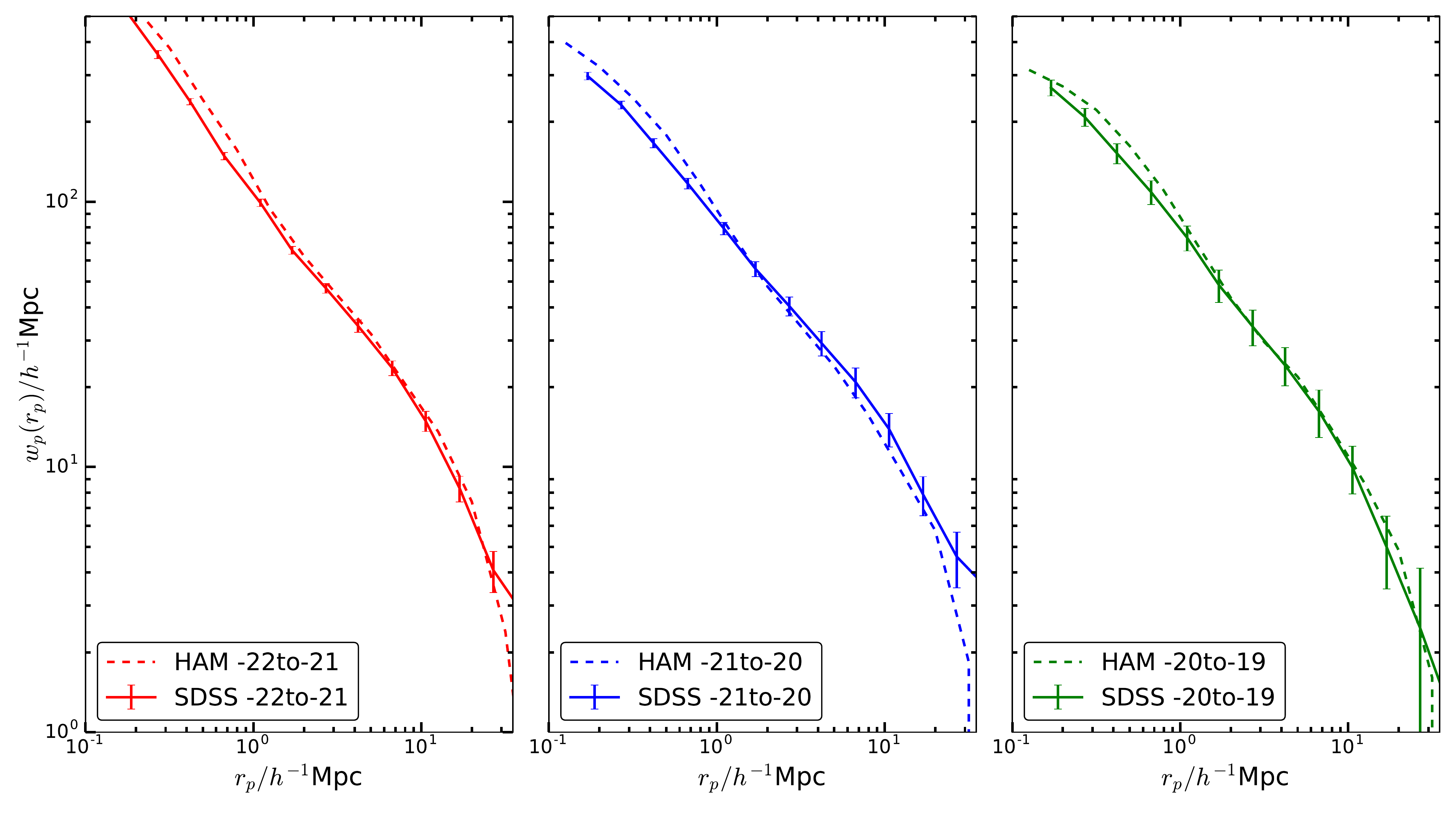}
\caption{Comparison of the projected 2-point correlation functions. The solid lines with error bars are extracted from \citep{zehavi2011galaxy}, and the dashed lines are given by our measurement on the HAM mock samples. The mock catalogue can reproduce the projected 2-point correlation functions from the observation for all three luminosity ranges.}\label{p2pcf}
\end{figure}

We follow the paper \citep{2pcfLS} to calculate the projected 2-point correlation function. Our results are shown in Fig.~\ref{p2pcf}. We have confirmed the previous results given by \citep{HAMmainref} that the HAM reproduced sample exhibits the same projected 2-point correlation functions as the observation in the luminosity range [-20,-19], [-21,-20] and [-22,-21], with differences mostly within 1 or 2 $\sigma$. We take these samples for further percolation analysis comparison. We shall also notice that the bias of the projected 2-point correlation function depends on the luminosity of the galaxy. We shall have a different view of the bias in the $S-bb$ relation.

\section{Result}
\subsection{$S-bb$ Relation Evolution}
\begin{figure}
\includegraphics[width=\textwidth]{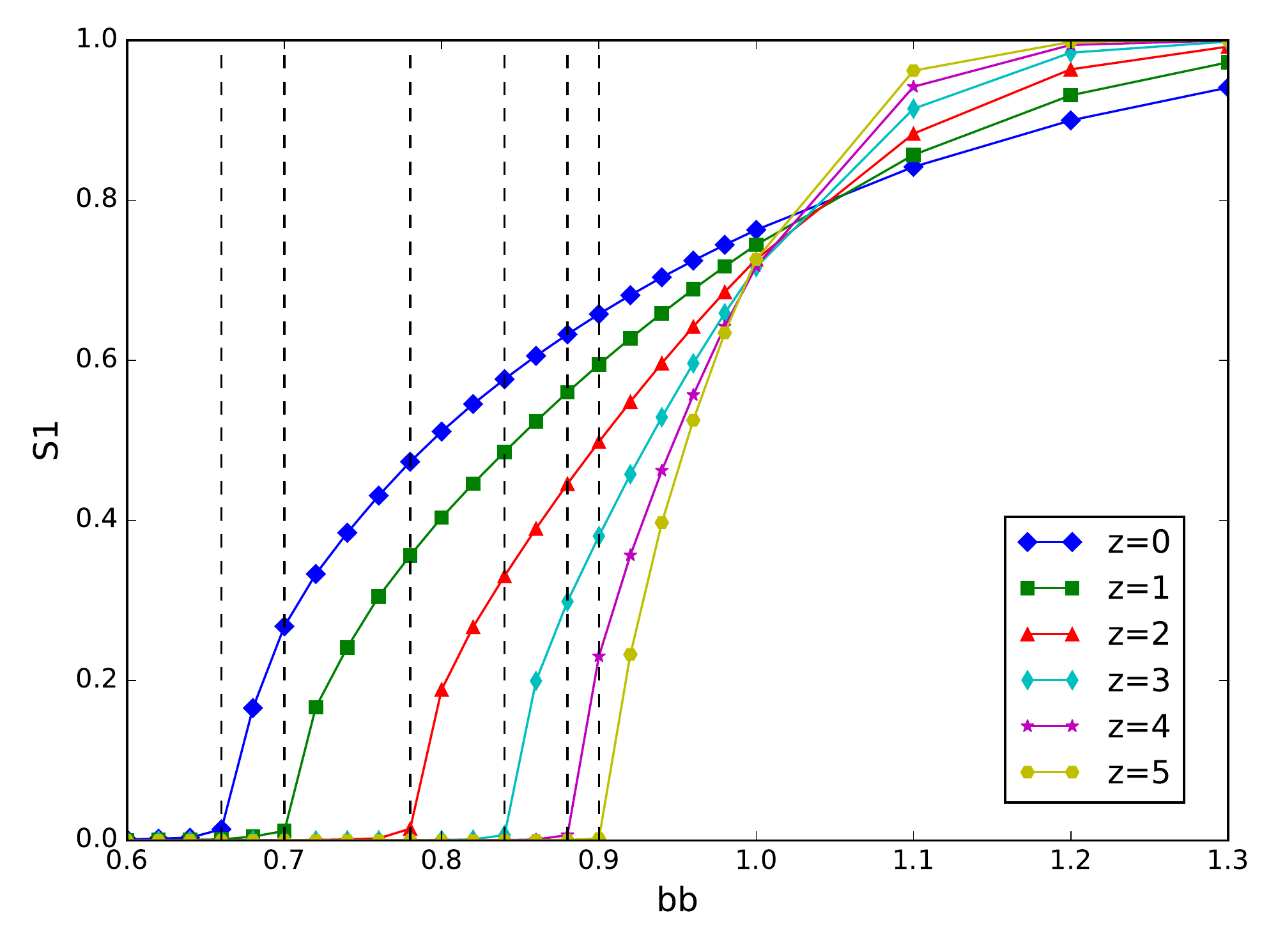}
\caption{The redshift ($z$) dependence of the $S-bb$ relation for $\Lambda$CDM cosmological simulation with $512^{3}$ particles in an $1h^{-1}$Gpc box. We can see that the transition threshold $bb_c$ gets smaller
for lower redshift, which is expected because the 2-point correlation function also becomes larger.}\label{S-bb-evolution}
\end{figure}

We are interested in the evolution of the cosmic web, which we study through an $\Lambda$CDM cosmological simulation with $512^{3}$ particles in a box with $1h^{-1}$Gpc sides, initial conditions generated by BBKS power spectrum \citep{BBKS} and Zel'dovich approximation \citep{ZeldovichApproximation}. We apply $\Omega_{m}=0.28$, $\Omega_{\Lambda}=0.72$ and $h=0.7$ for this test. As shown in Fig.~\ref{S-bb-evolution}, the $S-bb$ relation at high redshifts looks similar to that of random distribution (Fig.~\ref{random}), while as the density contrast increases with time, the $S-bb$ relation at low redshift becomes rather different, which has already been known in many previous works \citep{PerSim1985dead,PerSim1996}. 

\subsection{Bias Determination}
\begin{figure}
\includegraphics[width=0.45\textwidth]{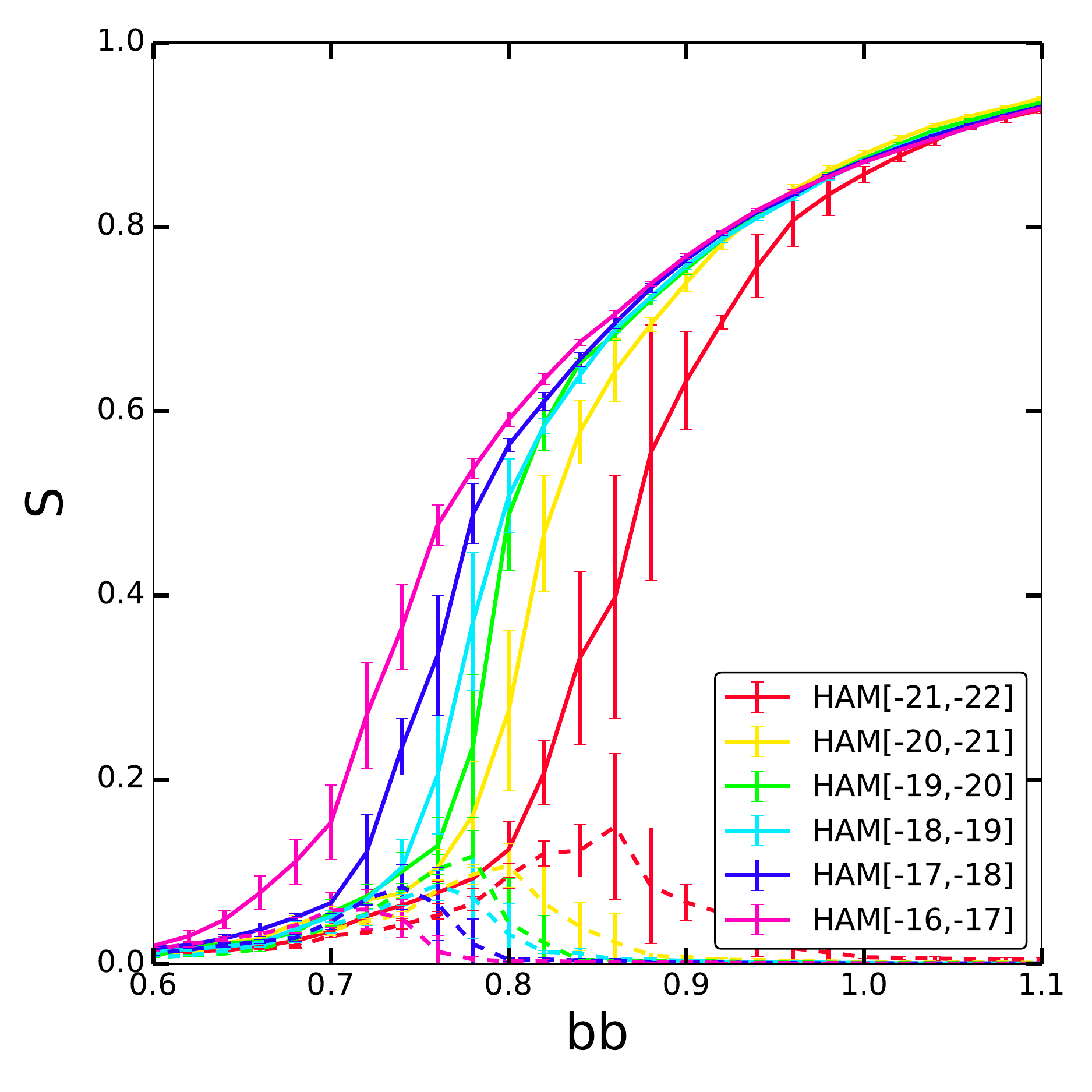}
\includegraphics[width=0.45\textwidth]{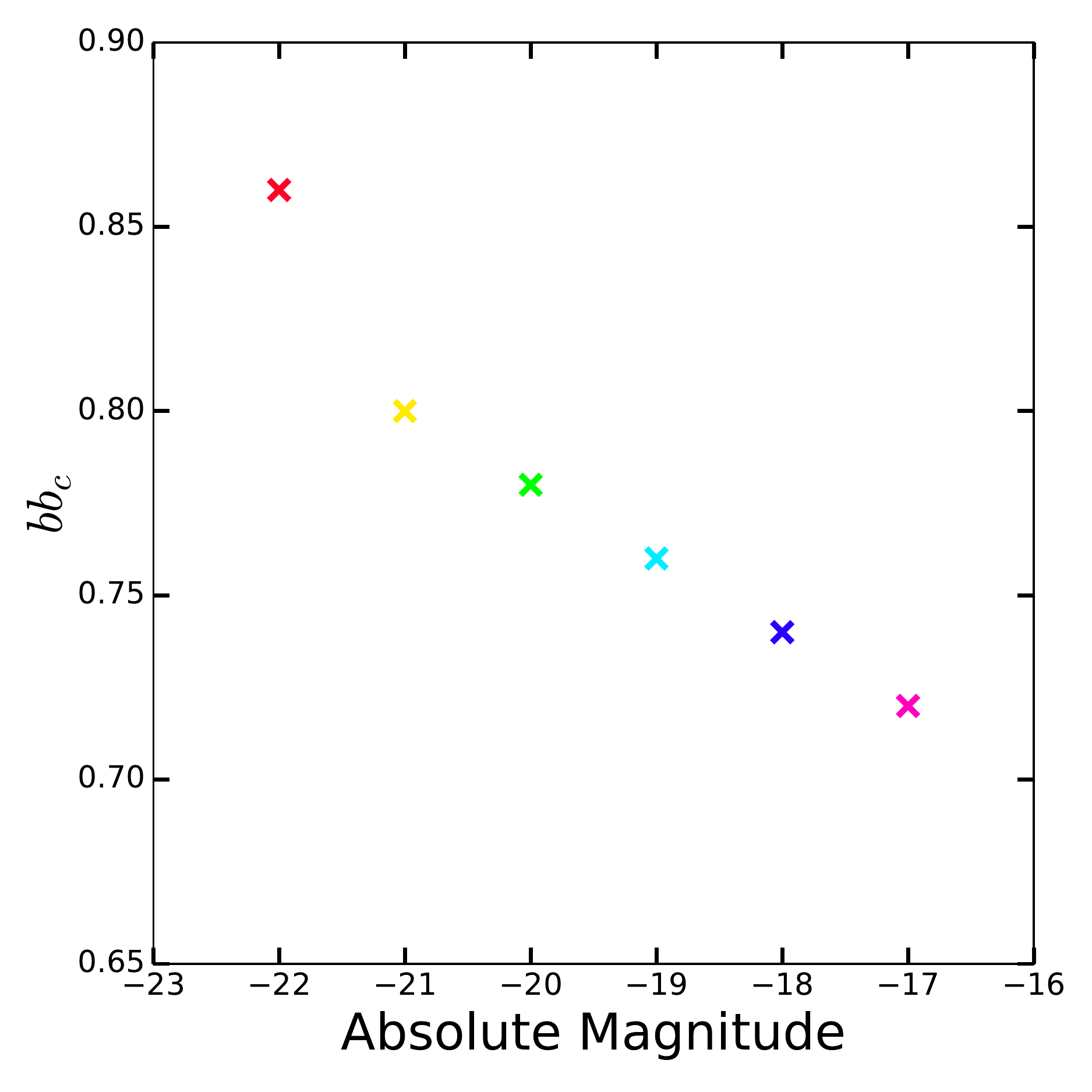}
\caption{The $S-bb$ relations of HAM mock samples with different luminosity range are shown here. From left to right, the lines are successively $S1-bb$ relation and the corresponding $S2-bb$ relation of HAM Mock samples of $-17<AM<-16$,$-18<AM<-17$,$-19<AM<-18$,$-20<AM<-19$,$-21<AM<-20$,$-22<AM<-21$. We  find that the lower luminosity samples' transition $bb$ is smaller than those of the high luminosity samples.}\label{S-bb-bias}
\end{figure}

The error bars of the $S-bb$ relation shown in fig.~\ref{S-bb-bias} are determined as follows. Based on Bolshoi simulation\citep{Bolshoi1}, we have only one realization of the HAM mock sample. So we use jackknife re-sampling to estimate the error bars. Here are the steps:
\begin{enumerate}
\item randomly choose $90\%$ of the HAM mock sample for ten times, generating 10 different mock samples.
\item Measure the $S-bb$ relations of these 10 mock samples.
\item Take the mean and standard variance of these 10 $S-bb$ relations as the value and error bar.
\end{enumerate}

\begin{figure}
\includegraphics[width=0.45\textwidth]{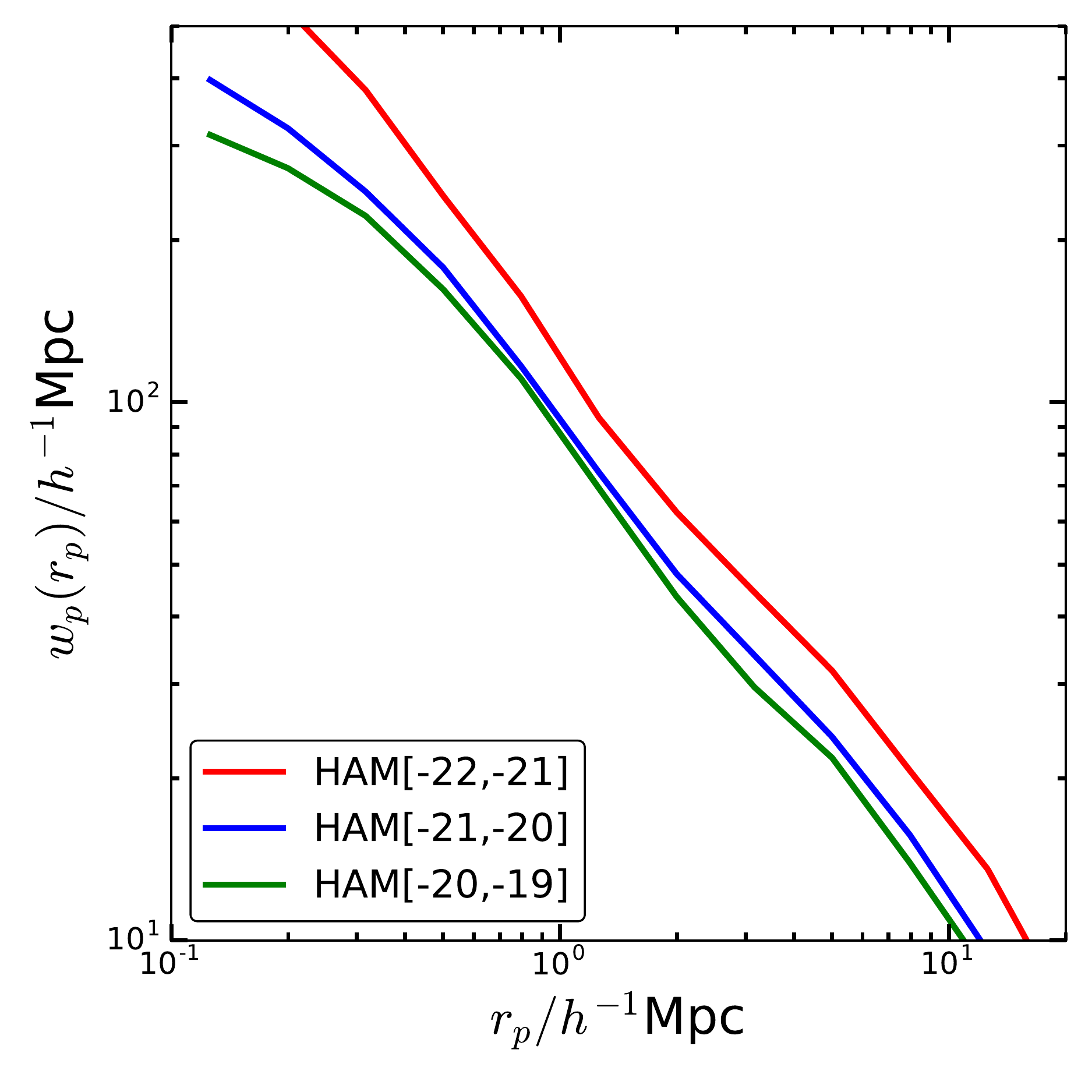}
\includegraphics[width=0.45\textwidth]{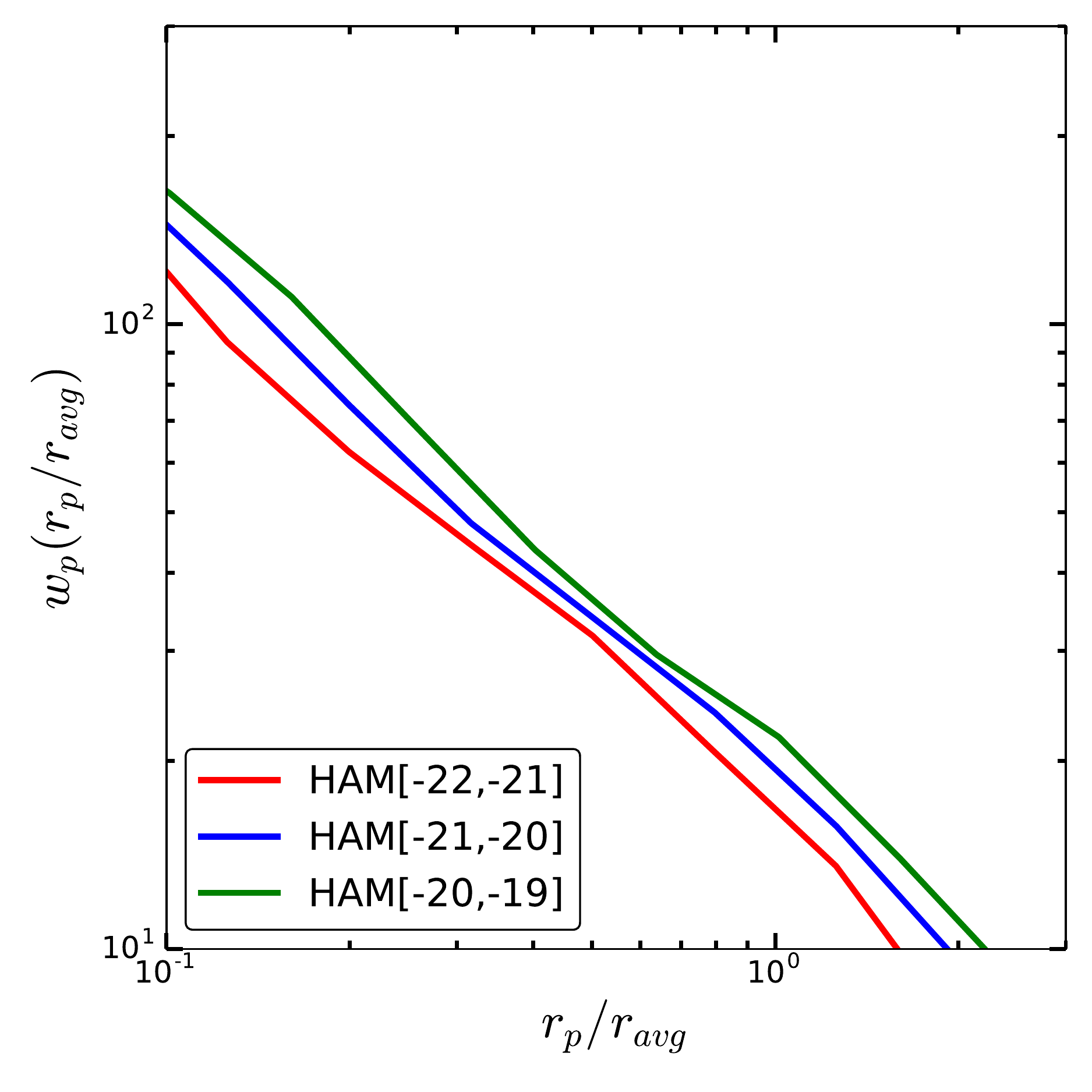}
\caption{Comparison between 2-point correlation functions normalized (right) and unnormalized (left) by the point average distance. Lower luminosity samples have higher normalized 2-point correlation function, but they have lower unnormalized 2-point correlation functions, which is well-known. Note that $r_{p}$ is the projected distance between a pair and $r_{avg}$ is the average distance of the points.}\label{normal-bias}
\end{figure}

From Fig.~\ref{S-bb-bias} we can see that the lower luminosity samples' transition threshold $bb$ is lower than those of the high luminosity samples. Based on this result, we may guess that the lower luminosity samples have higher bias in the 2-point correlation function than higher luminosity samples. However it is well known that more luminous galaxies have higher bias which seems to be contradictory to our results here. Note that the definition of $bb$ is such that it is normalized to the average distance of points. So we shall look at the 2-point correlation function with the projected distances $r_p$ also normalized to the average distance, which is shown in Fig.~\ref{normal-bias}. The higher the luminosity of the samples are, 
the lower their normalized 2-point correlation function is. Therefore, a smaller transition threshold is well
expected.

\begin{figure}
\includegraphics[width=0.45\textwidth]{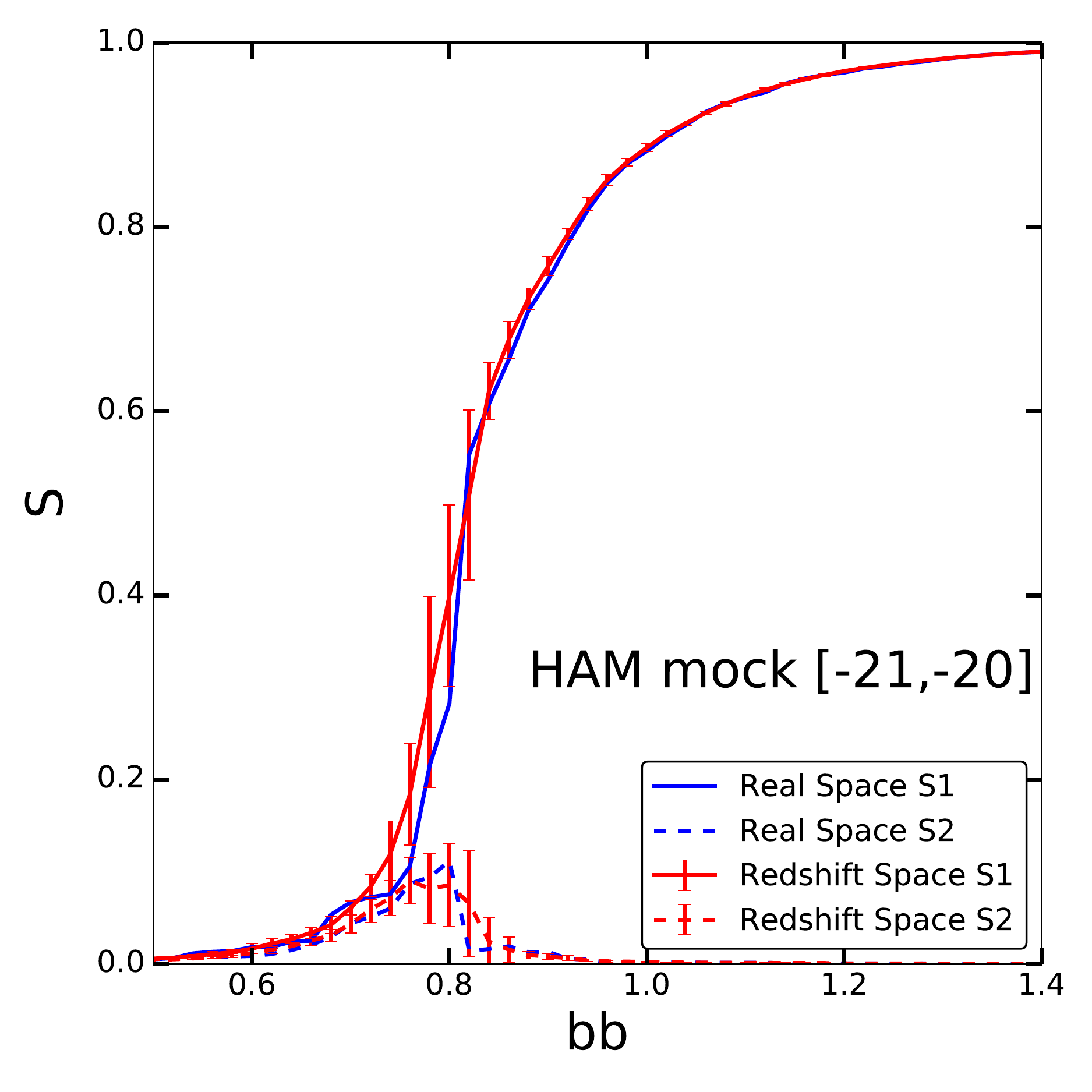}
\includegraphics[width=0.45\textwidth]{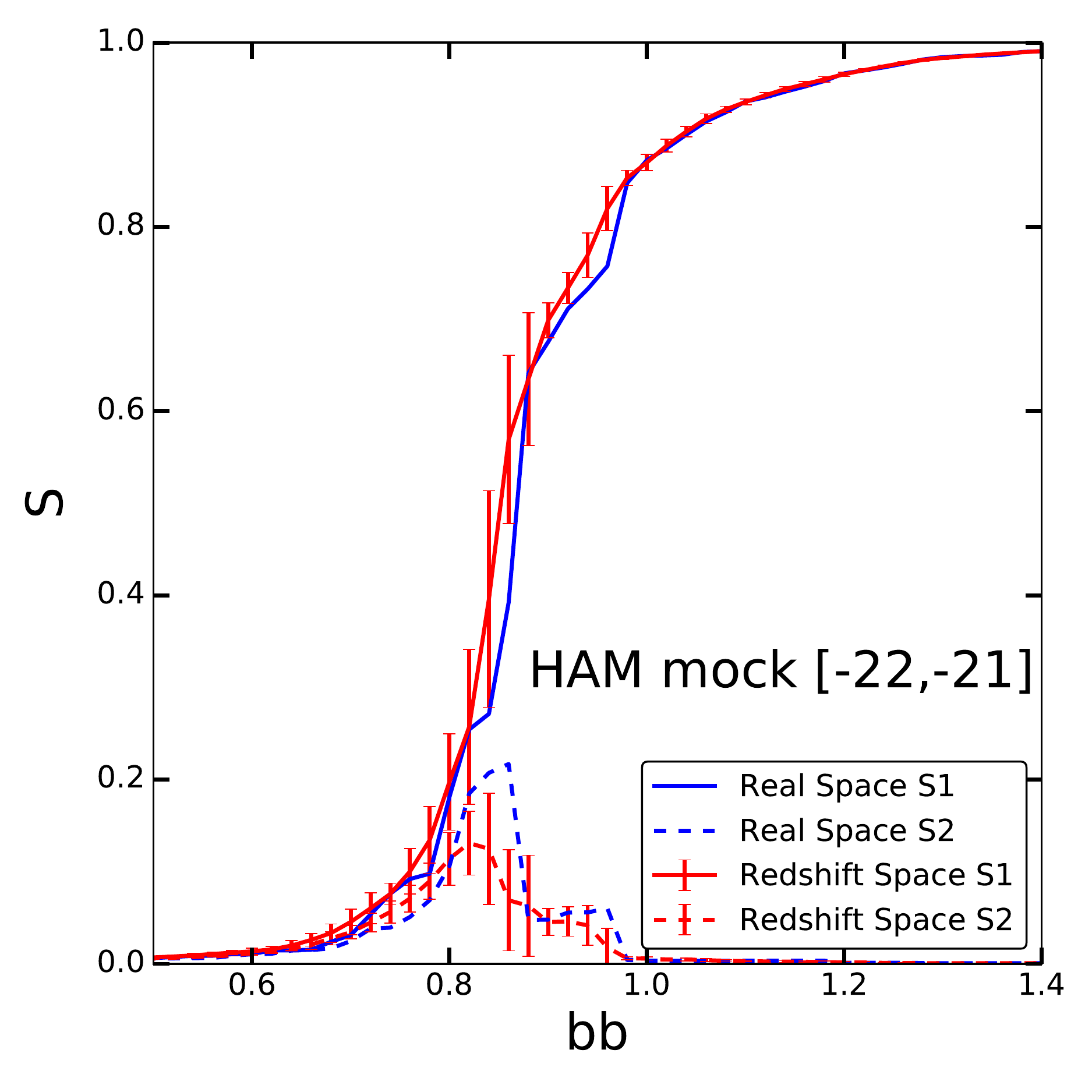}
\caption{The $S-bb$ relations of the HAM mock samples in real space (curve without error bars) and in redshift space (curves with error bars). The error bars are given by the standard variance of looking at the same sample from 14 different directions. The center of the sample is at redshift $z=0.1$. 
Clearly, the $S-bb$ relation is not sensitive to redshift distortion.}\label{redshift-test}
\end{figure}

Galaxies' real spatial distribution is different from their redshift space distribution due to the peculiar velocities of the galaxies. Using the HAM mock sample data with peculiar velocity, we can study the redshift distortion effect in the $S-bb$ relation. We uniformly choose 14 different directions to look at the sample, all of which are about $z=0.1$ away from the sample center. The distance is translated by the cosmological parameters used in the Bolshoi simulation. As shown in Fig.~\ref{redshift-test}, we find that redshift distortion has little effects on the $S-bb$ relation. Because the $S-bb$ relation shows mainly the topological information of the cosmic web, it is reasonable that it is not sensitive to local distortions. 
\begin{figure}
\includegraphics[width=\textwidth]{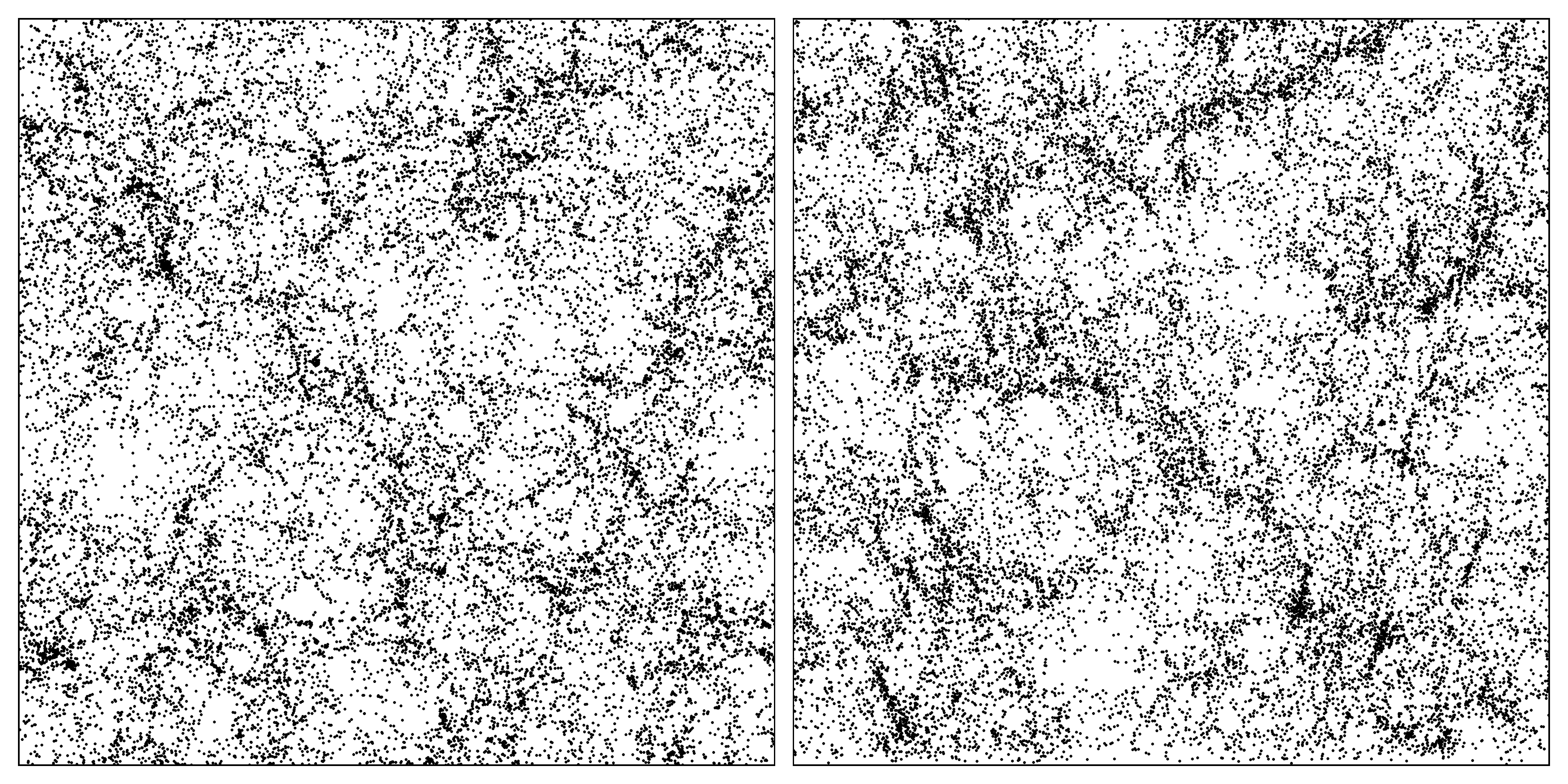}
\caption{A slice of galaxy distribution in real space (on the left) and the same slice in
redshift space (on the right) as an illustration to show the effect of redshift distortion. The slice
is $100h^{-1}$Mpc in thickness, $250h^{-1}$Mpc in length and width, taken from the mock catalogues we 
generated. These galaxies' absolute magnitudes are between $-20$ and $-21$.}
\end{figure}

Our observation data comes from SDSS DR12 \citep{SDSSDR12}. We take the North Galactic Cap (NGC) as our galaxy zoo region because of the continuous, large coverage of the NGC. Thus we have the spectrum of galaxies covering huge volume in the red-shift space in NGC. We obtain the data of galaxies in the range of $130<RA<230$, $10<Dec<50$, and $0.07<z<0.14$, including their coordinates, spectral redshifts and Petrosian magnitudes in the $r$ band. The volume of this region is quite similar with the volume of the Bolshoi simulation. Then we use the same cosmological parameters as the Bolshoi simulation to calculate the luminosity distances of the galaxies using their spectral redshifts. We also calculate the Petrosian absolute magnitudes of the galaxies. But we didn't correct their absolute magnitudes. So the magnitudes of the galaxies are in fact in the $r^{0.1}$ band as described in \citep{LumFun}, since the Luminosity Function we use to generate the HAM mock galaxy catalogue is also in the $r^{0.1}$ band. The K-correction for this sample is also not important because we have chosen the red shift range to be $0.07<z<0.14$. According to the discussion in \citep{Kcorrection}, the K-correction for $r^{0.1}$ is no more than 0.1. So we just ignore K-correction for our observation sample. Now we have generated an observation galaxy catalogue with 3-D space positions and reasonable absolute magnitudes. The volume covered by the sample is also similar to the volume covered by the HAM mock sample.

Before SDSS DR12, there was no such large continuous angular coverage as well as deep redshift coverage. The public access to the SDSS DR12 allows us to play with a unprecedented, continuous, large scale, 3-D galaxy zoo. It is a good time to search for unusual statistical measurement of the large scale structures, not just the percolation analysis we proposed here but also many other ways of viewing the structures as a whole web.

However, we have to point out that the observational sample is in redshift space while the HAM mock sample is in real space. So the direct comparison of these two samples is not fair. We must take redshift distortion into consideration. Fortunately, our HAM mock sample also contains information of peculiar velocity, which means we can easily transfer the HAM mock sample into the redshift space.

\begin{figure}
\includegraphics[width=\textwidth]{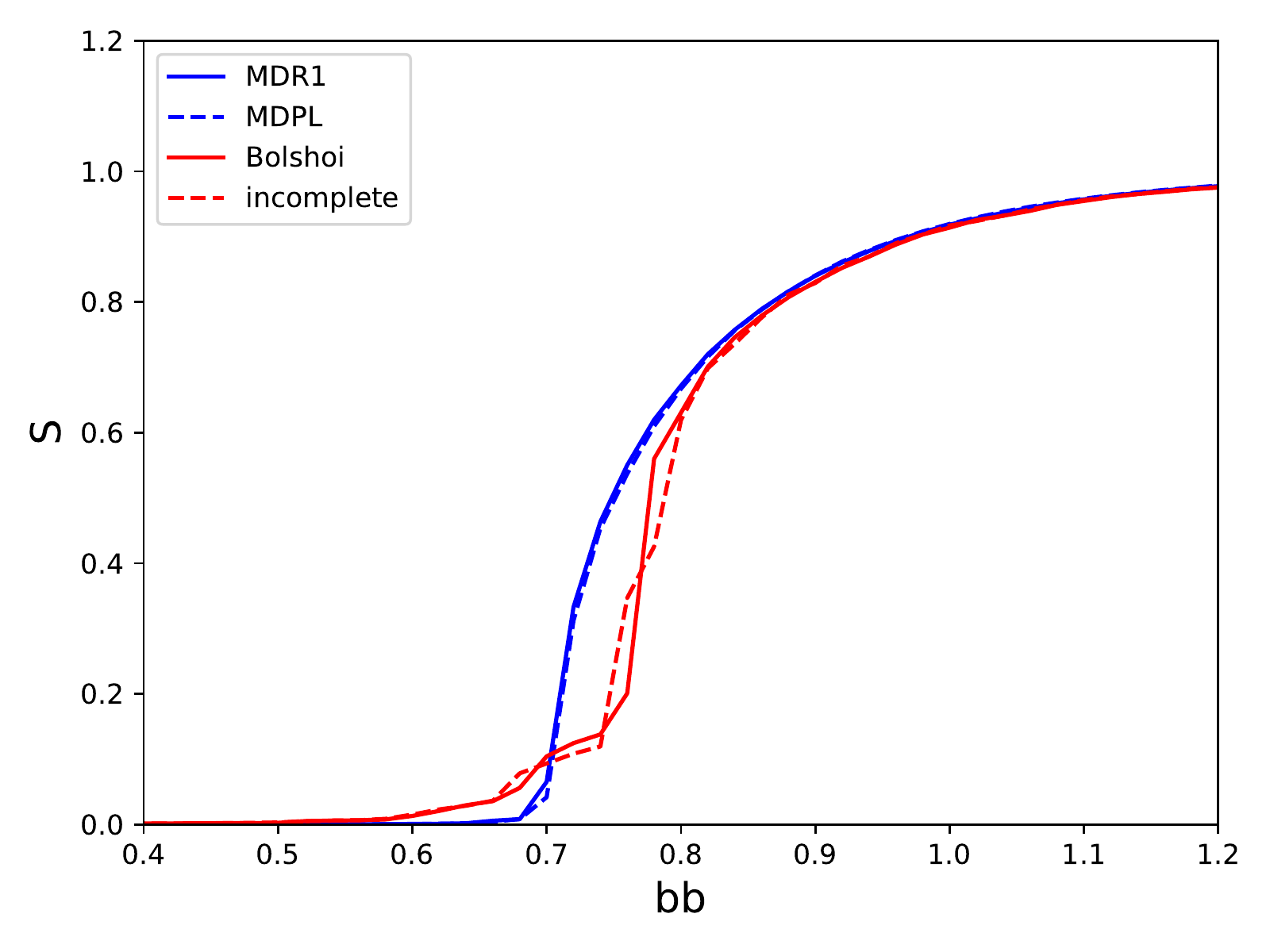}
\caption{$S_1-bb$ relation or simply $S-bb$ relation measured from the halo catalogue with halo mass $>10^{12}M_{\odot}$ built from MDR1 simulation \citep{multidark1}(WMAP5 cosmology), MDPL simulation \citep{multidarkplanck}(Planck cosmology) and Bolshoi simulation \citep{Bolshoi1}(WMAP5 cosmology). The red dashed line is measured from the incomplete sample generated from the Bolshoi sample considering fiber collision. The different cosmology between MDR1, Bolshoi and MDPL, and the incompleteness due to fiber collision affect the $S-bb$ relation little. However, the larger boxsizes of the MDR1 and the MDPL simulations introduce sharper transition in the $S-bb$ relation comparing to the Bolshoi simulation.}\label{fiber and planck}

\end{figure}

We also need to consider the incompleteness of the observational sample. As shown in the DR12 paper \citep{SDSSDR12}, that is mainly because of fiber collision. We can also simulate the incompleteness by excluding those haloes in the simulation that looks closer than $62"$ projected on the 2D plane due to fiber collision, which will exclude $\sim 5\%$ halos mainly in high density region. From Fig.~\ref{fiber and planck} we can see that, this incompleteness has negligible effect for the $S-bb$ relation.
Therefore, the $S-bb$ relation is very robust against the observational selection bias introduced by fiber collision. We also find that, changing from the adopted cosmological parameters in both Bolshoi and MDR1 simulations to Planck cosmology, the $S-bb$ relation will not be changed significantly. However, due to the limited boxsize of Bolshoi simulation, the difference between the $S-bb$ relation of MDR1 and Bolshoi simulations is observable. That will cause the misidentification of the transition threshold by $0.04$, but for $bb>0.8$, the $S-bb$ relation difference is negligible. Therefore, using the HAM mock sample built from the Bolshoi simulation to compare with observation is fair, but we need to be aware of a small misidentification of the transition threshold, because of the limited size of the simulation.

\subsection{Comparison with SDSS DR12}
\begin{figure}
\includegraphics[width=0.45\textwidth]{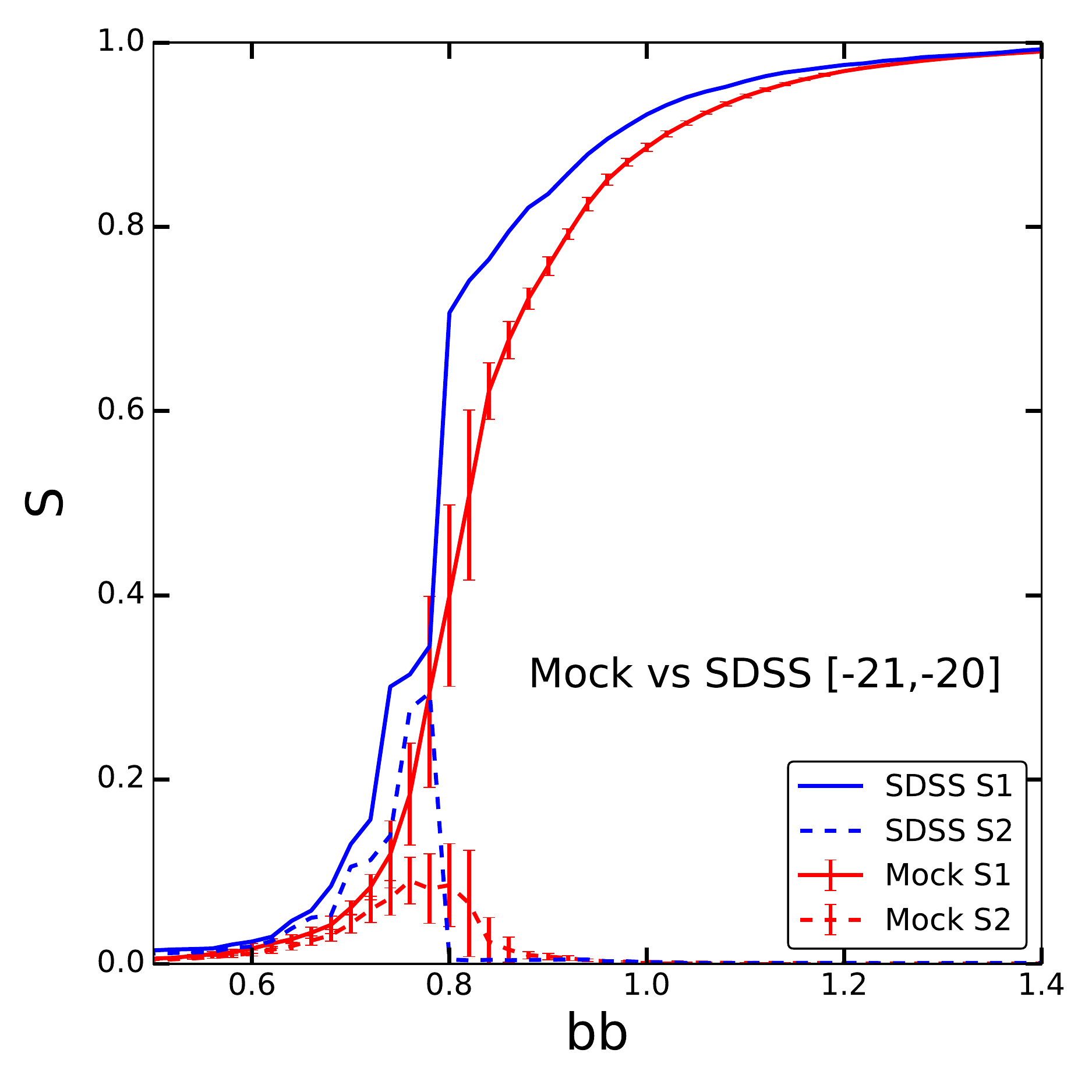}
\includegraphics[width=0.45\textwidth]{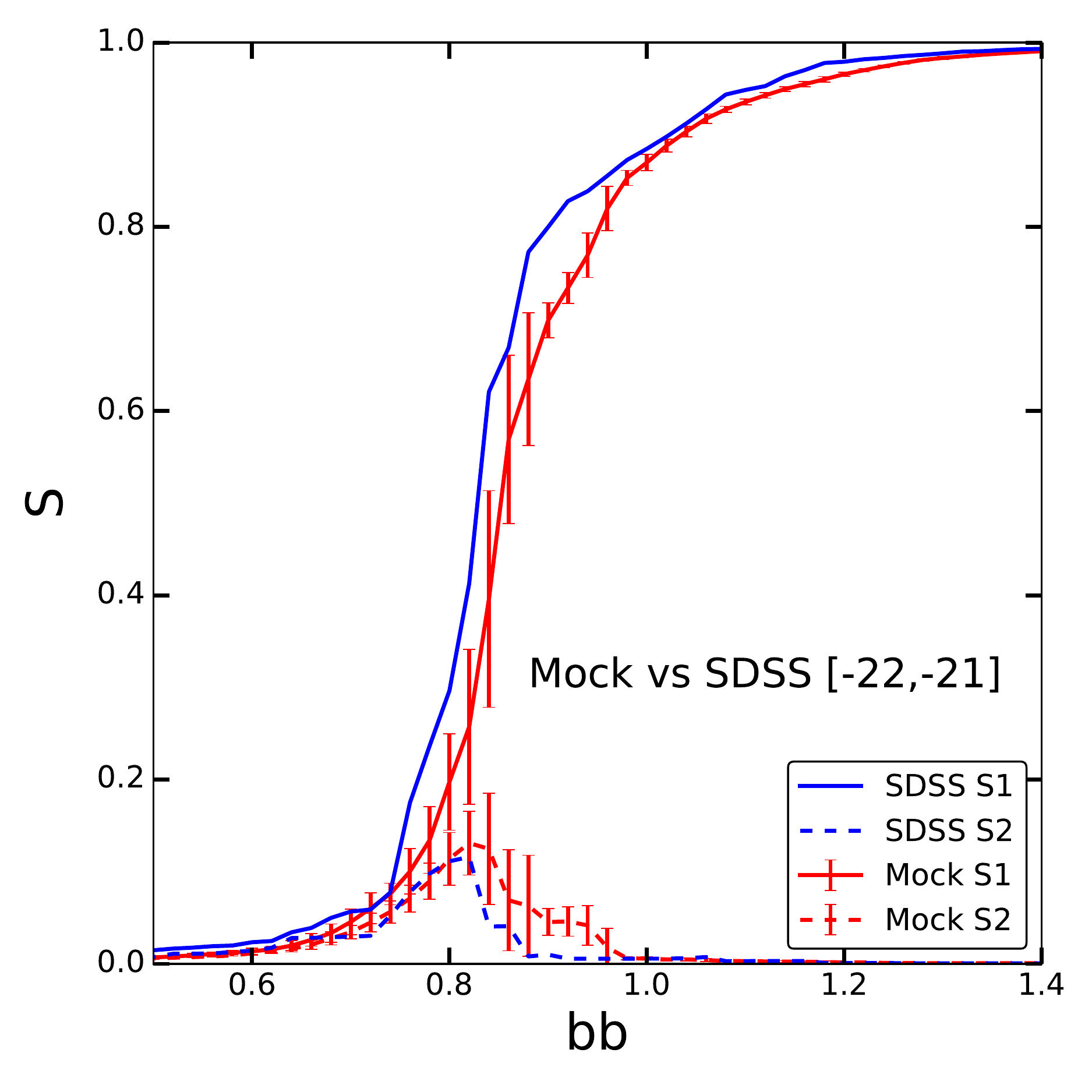}
\caption{Comparison of the $S-bb$ relations from SDSS DR12 galaxy sample (blue curves) and the HAM mock sample (red curves). The corresponding $S_{2}-bb$ relations are plotted in dashed curves in order to show the transition thresholds. The $S-bb$ relation is different for the mock sample and observation, which is hard to be explained by redshift distortion, incompleteness of observation or different cosmological parameters (Planck and WMAP5).}\label{S-bb-compare}
\end{figure}
Here we are comparing the observational galaxy sample and the HAM mock galaxy sample in two luminosity ranges $-22<M<-21$ and $-21<M<-20$, as shown in Fig.~\ref{S-bb-compare}. We choose these two ranges because, firstly, we need generally enough number of galaxies to reduce the Poisson error and secondly, we require the galaxies to be detectable even at red shift $z=0.14$ to get rid of Malmquist bias.  Interestingly, although the projected 2-point correlation functions of the HAM mock sample and observed sample are quite similar, their $S-bb$ relations are different by 2-3 $\sigma$, particularly the transition point in the $S_{2}-bb$ relation. According to our PCCET, this is most probably because the HAM mock sample doesn't reproduce the high order correlation functions well enough. This is also supported by a recent work \citep{Hong2016Net} comparing Illustris simulation and the data sample generated with the same two-point correlation function. The measured $S-bb$ relations (called 'giant component-linking length' relation), for the two samples are significantly different. The transition threshold is also smaller for the real simulation sample. 

\section{Conclusion and Discussion}
We have proposed the $S-bb$ relation as a branch of percolation analysis of the cosmic web. We have shown that the $S-bb$ relation and its transition threshold reveal physical information of the cosmic web. It also has some advantages compared to the standard 2-point (n-point) correlation function:
\begin{enumerate}
\item The $S-bb$ relation is based on the Friend-of-Friend (FoF) algorithm, which is easy to implement and costs very little computational resources.
\item Our theoretical model, PCCET, shows that the $S-bb$ relation contains information of all orders of correlation functions.
\item The $S-bb$ relation is robust against redshift distortion. 
\item The $S-bb$ relation is robust against incompleteness, which is very useful in observation. We have shown that $90\%$ completeness is quite acceptable and this has been reached in SDSS.
\item The $S-bb$ relation gives information on the bias of the correlation function.  
\end{enumerate}
However, it also has some limitations:
\begin{enumerate}
\item We need large sample of galaxies to look at the $S-bb$ relation, with at least $100h^{-1}$Mpc box side length.
\item We need large continuous region of observation to study the $S-bb$ relation, especially for 3-D analysis.
\item It is not easy to extract every order of the correlation functions from the $S-bb$ relation.
\end{enumerate}
Cellular or clumpy? That's not the question any more. It is well known that the distribution of galaxies is cellular. How cellular is the cosmic web? That's the question. We suggest the $S-bb$ relation as a new measurement, which is applicable to a wide range of data types from simulation particles to galaxy catalogue, fast to compute (as fast as 2-point correlation function measurement), robust (against redshift distortion and incompleteness), and it contains information of all orders of correlation functions. In the framework of PCCET, we have good understanding of the $S-bb$ relation. We have compared the $S-bb$ relation of the HAM mock galaxy catalogue with SDSS DR12 galaxy catalogue. A significant difference was found, though these two samples had similar projected 2-point correlation functions. The percolation analysis with discrete points can help us understand the cosmic web. At the minimal, it provides a quick check of whether our mock sample really represents the real distribution or not. 

\section{Appendix}
\subsection{Proof for PCCET}\label{sec:SbbApp}
The series in Eq.~\ref{eq:S-P} is known q-Pochhammer function or Euler function in mathematics.
The q-Pochhammer symbol $(a;q)_n$ is defined as 
\begin{equation}
\begin{split}
                (a;q)_n & = \prod_{k = 0}^{n-1}(1-aq^k)   \\
                        & = (1-a)(1-aq)(1-aq^2)\cdots(1-aq^{n-1}).
\end{split}
\end{equation}
In particular, the Euler function $\phi(q)$ is defined as the infinite product
\begin{equation}
\phi(q) = (q;q)_{\infty}, 
\end{equation}
\begin{equation}
q = 1 - p.
\end{equation}                             
The $S_n(p)$ can be expressed in terms of the q-Pochhammer symbols 
\begin{equation}
S_n(p) = \frac{1}{n}\sum_{k=0}^{n-1}(q;q)_k,
\end{equation}
where
\begin{equation}
0 < q < 1.
\end{equation}
Since $0<q<1$, it is not difficult to realise the fact that
\begin{equation}
 1 > (q;q)_1 > (q;q)_2 > \cdots (q;q)_n > 0.
\end{equation}                        
Therefore
\begin{equation}
 S_n(q) > \frac{1}{n}n(q;q)_{n-1} = (q;q)_{n-1}.
\end{equation}
On the other hand,
\begin{equation}
S_n(q) = \frac{1}{n}\sum_{k = 0}^{n - 1}(q;q)_{k} = \frac{1}{n}\sum_{k = 0}^{m - 1}(q;q)_k + \frac{1}{n}\sum_{k = m}^{n - 1}(q;q)_k.
\end{equation}
Here we introduce an integer $m$ such that
\begin{equation}
                            m = INT(\sqrt{n}), 
\end{equation}
where $INT(x)$ is defined as
\begin{equation}
INT(x) = max\{n \in Z | n \leq x\}.
\end{equation}
Since as $n$ increases, $(q;q)_n$ decreases, and we have 
\begin{equation}
\begin{split}
        S_n(q) & = \frac{1}{n}\sum_{k = 0}^{m - 1}(q;q)_k + \frac{1}{n}\sum_{k = m}^{n - 1}(q;q)_k  \\
             & < \frac{m}{n}(q;q)_0 + \frac{n - m}{n}(q;q)_m \\
             & = \frac{m}{n} + \frac{n - m}{n}(q;q)_m.
\end{split}
\end{equation}
In summary,
\begin{equation}
(q;q)_{n-1} < S(n) < \frac{m}{n} + \frac{n - m}{n}(q;q)_m.
\end{equation}
$m$ as a function of $n$ has the following asymptotic properties
\begin{equation}
\begin{split}          
                      & \lim_{n \rightarrow \infty} m = \infty,   \\
                      & \lim_{n \rightarrow \infty} \frac{m}{n} = 0.
\end{split}
\end{equation}
For simplicity, we denote
\begin{equation}
L_n(q) = (q;q)_{n-1},  \qquad U_n(q) = \frac{m}{n} + \frac{n - m}{n}(q;q)_m.
\end{equation}
Then 
\begin{equation}
\begin{split}
\lim_{n \rightarrow \infty} L_n(q) = (q;q)_{\infty} = \phi(q), \\
\lim_{n \rightarrow \infty} U_n(q) = (q;q)_{\infty} = \phi(q).
\end{split}
\end{equation}
Since we always have
\begin{equation}
L_n(q) < S_n(q) < U_n(q),
\end{equation}
according to the Squeeze Theorem, we concluded that
\begin{equation}
\lim_{n \rightarrow \infty} S_n(q) = \phi(q).
\end{equation}
In the form of a step function, we find the limit of $S_n(p)$ to be:
\begin{equation}
         \lim_{n \rightarrow \infty} S_n(p) = 
          \begin{cases}
                0 & \text{if } p = 0, \\
               \phi(1-p) & \text{if } 0 < p < 1, \\
                1 & \text{if } p = 1.
          \end{cases}
\end{equation}
\section*{Acknowledgements}

J. Zhang thanks Prof. Y. Jing for giving quite useful advices, L. Yang and Z. Li for helping with measuring 2-point correlation function, Dylan for useful discussion, S. Liao for useful discussion and advice throughout this project, J. Chen for his great help in studying the PCCET mathematically in depth. 

“The CosmoSim database used in this paper is a service by the Leibniz-Institute for Astrophysics Potsdam (AIP).
The MultiDark database was developed in cooperation with the Spanish MultiDark Consolider Project CSD2009-00064.”
    
    The authors gratefully acknowledge the Gauss Centre for Supercomputing e.V. (www.gauss-centre.eu) and the Partnership for Advanced Supercomputing in Europe (PRACE, www.prace-ri.eu) for funding the MultiDark simulation project by providing computing time on the GCS Supercomputer SuperMUC at Leibniz Supercomputing Centre (LRZ, www.lrz.de).

    The Bolshoi simulations have been performed within the Bolshoi project of the University of California High-Performance AstroComputing Center (UC-HiPACC) and were run at the NASA Ames Research Center.

\bibliography{sbb}
\end{document}